\pdfoutput=1

\documentclass[letterpaper, 10 pt, conference]{ieeeconf}
\IEEEoverridecommandlockouts
\usepackage[utf8]{inputenc}
\usepackage[T1]{fontenc}
\usepackage{amsmath,amssymb}
\usepackage{bm}
\usepackage{graphicx}
\usepackage{booktabs}
\usepackage{siunitx}
\usepackage{cite}
\usepackage{url}

\usepackage{pgf}
\usepackage{tikz}
\usepackage{pgfplots}
\pgfplotsset{compat=1.18}

\newtheorem{remarkx}{Case}
\newenvironment{remark}[1][]{\begin{remarkx}[#1]\normalfont}{\end{remarkx}}

\usepackage{eso-pic}

\newif\ifarxiv
\arxivtrue

\newcommand{\arxivnotice}{%
  This work has been submitted to the 2027 European Control Conference (ECC)
  for possible publication. Copyright may be transferred without notice,
  after which this version may no longer be accessible.}

\begin{document}

\title{%
Data-Driven Predictive Control for Autonomous Vehicle Racing under Varying Friction
}

\author{Julius Beerwerth$^{1}$, Julius Tischbein$^{2}$, and Bassam Alrifaee$^{1}$%
\thanks{Code, simulation data, and the hardware evaluation runs are available at \protect\url{https://github.com/bassamlab/multisurface-racing-with-dpc}.}%
\thanks{$^{1}$J.~Beerwerth and B.~Alrifaee are with the Department of Aerospace Engineering, University of the Bundeswehr Munich, 85577 Neubiberg, Germany. {\tt\small \{julius.beerwerth, bassam.alrifaee\}@unibw.de}}%
\thanks{$^{2}$J.~Tischbein is with the Department of Computer Science, RWTH Aachen University, 52062 Aachen, Germany. {\tt\small julius.tischbein@rwth-aachen.de}}%
}

\maketitle

\ifarxiv
  \AddToShipoutPictureBG*{%
    \AtPageLowerLeft{%
      \raisebox{1.5\baselineskip}{%
        \makebox[\paperwidth]{%
          \parbox{0.85\paperwidth}{\centering\footnotesize\arxivnotice}}}}}
\fi

\begin{abstract}
Data-driven predictive control (DPC) predicts future outputs by combining recorded input--output trajectory data such that the combined trajectory approximately matches a measured initial trajectory and tracks a provided reference. This paper investigates to what extent and under which conditions DPC using data pooled across operating conditions retains the prediction accuracy of DPC using data from the current operating condition.
In autonomous vehicle racing, the tire--road friction varies between racetrack sections and tire sets. 
If the initial trajectory and the recorded data originate from different frictions, prediction accuracy of DPC degrades. Ensuring the frictions match requires knowing the friction during recording and estimating the current friction explicitly.
Instead, we pool data recorded at different frictions. We compare the prediction accuracy of DPC using data recorded under the current friction (matched), a different friction (mismatched) and data pooled from both frictions. We evaluate in simulation and small-scale vehicle experiments in the Cyber-Physical Mobility Lab (CPM Lab).
Our results show that pooled data keep the multi-step root-mean-square prediction error within 1.5 times that of matched data, whereas mismatched data increase it by a factor of 2.2 to 3.4.
To assess how this prediction accuracy carries over to control, we evaluate the tracking error from the raceline in closed loop.
In simulation, DPC with pooled data tracks comparably to nonlinear model predictive control (\qty{13.7}{\centi\metre} versus \qty{13.3}{\centi\metre}) and close to DPC with matched data in the CPM Lab, differing by \qty{0.1}{\centi\metre} on the low-friction tires (\qty{1.6}{\centi\metre} versus \qty{1.5}{\centi\metre}) and by \qty{0.2}{\centi\metre} on the high-friction tires (\qty{0.9}{\centi\metre} versus \qty{1.1}{\centi\metre}).
DPC thus provides partial adaptation to varying friction, without a dedicated adaptation method.
\end{abstract}

\section{Introduction}\label{sec:intro}

\subsection{Motivation}

Data-driven predictive control (DPC) predicts future outputs by combining recorded input--output trajectory data such that the combined trajectory approximately matches a measured initial trajectory and tracks a provided reference \cite{coulson2019DataEnabled,berberich2025Overview}. For linear systems, a linear combination of recorded trajectories is itself a valid trajectory of the system. For nonlinear systems, a linear combination of recorded trajectories only approximates the system dynamics. Moreover, recorded trajectories capture the system dynamics only at the operating conditions at which they were collected. Data recorded at an operating condition different from the current one may degrade the prediction accuracy of DPC. Ensuring that the recorded data match the current operating condition requires knowing the operating condition during recording and estimating the current one.

In autonomous vehicle racing such variations in operating conditions are particularly relevant. The tire--road friction, hereafter friction, varies between racetrack sections and tire sets. We refer to specific tire--road frictions as friction levels. Each recorded trajectory only captures the vehicle dynamics under the friction level it was recorded at. Using data recorded at a friction level different from the current one degrades the prediction accuracy of DPC.

Controllers for autonomous vehicle racing are commonly based on model predictive control (MPC), which predicts vehicle motion using an identified vehicle model \cite{betz2023TUM,scheffe2023Sequential}. The tire parameters of the vehicle model are influenced by friction, so the accuracy of a vehicle model identified at one friction level degrades at other friction levels \cite{laurense2017Pathtracking}. To retain performance under varying friction, many model-based approaches adapt the vehicle model online. Such online adaptation requires structural model knowledge and a dedicated adaptation method \cite{vaskov2023Frictionadaptive,kalaria2024Adaptive}.

Instead of explicitly identifying the current friction, we pool data recorded at different friction levels. Since the measured initial trajectory can contain information about the current friction, DPC with pooled data may favor recorded trajectories that match it. However, trajectories that do not match the current friction can still enter the combined trajectory and degrade its prediction accuracy. This paper investigates to what extent and under which conditions DPC using data pooled across friction levels retains the prediction accuracy of DPC using data from the current friction level. %

\subsection{Related Work}

\textit{Model-based control under varying friction.}
Model-based controllers for autonomous vehicle racing describe the tire forces with empirical tire models such as the Pacejka model \cite{zarrouki2024Stochastic}. To account for model mismatch, Gaussian-process-augmented MPC learns a residual to a vehicle model from data \cite{hewing2020Cautious,kabzan2019Learningbased}, and Nagy et al.\ train one Gaussian process per friction and combine them online \cite{nagy2023Ensemble}. AL-Sunni et al.\ evaluate a set of vehicle models with different tire parameters in parallel and select the vehicle model that best explains the recent vehicle dynamics \cite{al-sunni2025LLAMPC}. Closest to our approach, Spielberg et al.\ train a neural network on data from dry and snow-covered roads, to predict the vehicle motion from a history of past states and inputs \cite{spielberg2019Neural,spielberg2022Neural}. In this paper, we neither select, adapt, nor train a vehicle model; instead, we apply DPC with data pooled from trajectories recorded at different friction levels.

\textit{DPC for nonlinear systems.}
Since vehicle dynamics are nonlinear, DPC's linear predictor only approximates the future vehicle motion. The prediction accuracy depends on which trajectories enter it. Existing methods update the trajectory data online \cite{berberich2022Linear,baros2022Online,zhou2025Effective,fayebedrin2026Structured},  select trajectories relevant to the current operating point \cite{naf2025Choose,reinhardt2023DataDriven,li2025Datamodel}, or penalize distant trajectories through a weighted regularization \cite{engeln2026Datadriven}. In our previous work, contextual sampling selects trajectories near the initial trajectory and reference \cite{beerwerth2025Less}. These methods measure relevance through the recorded inputs and outputs, but do not account for unmeasured variables, such as friction.

\textit{DPC across operating conditions.}
Klädtke and Schulze Darup, as well as Giacomelli et al., analyze which recorded trajectories DPC uses in its prediction \cite{kladtke2025Data,giacomelli2025Insights}; similarly, we analyze at which friction level the trajectories that DPC uses in its prediction were recorded.
If the system dynamics depend on a measured variable, DPC can organize the data by this variable. Linear parameter-varying DPC uses a measured scheduling signal to determine the current dynamics \cite{verhoek2021Datadriven,verhoek2026Linear}, and gain-scheduled DPC partitions the data into regions of a measured scheduling variable and switches between the regional datasets \cite{guerrero2025GainScheduled,zieglmeier2025GainScheduling}. Without a scheduling variable, Han et al.\ partition the data into linear subsystems and identify the relevant one from the current initial trajectory \cite{han2024Big}. In this paper, the friction could be such a scheduling variable, but friction is difficult to measure or estimate. 

Instead, we record data at the current friction level (matched data) and at another friction level (mismatched data) and pool them into one dataset. We investigate to what extent and under which conditions DPC with pooled data retains the prediction accuracy of DPC with matched data.

\subsection{Contributions}

The contributions of this paper are:
\begin{enumerate}
\item We decompose the prediction of DPC using data pooled from trajectories recorded at different friction levels into matched and mismatched contributions. Building on that, we propose practical conditions under which nonlinear DPC using pooled data predicts nearly as accurately as using matched data, without explicit friction estimation or a dedicated adaptation method.
\item We show in simulation that DPC using pooled data predicts nearly as accurately as using matched data, whereas mismatched data degrade the prediction accuracy. We introduce an implied friction and show that the pooled predictions predominantly resemble the matched predictions where friction influences the vehicle motion.
\item We show in simulation that DPC with pooled data achieves tracking performance comparable to the nonlinear model predictive control (NMPC) from the TUM-CONTROL framework \cite{zarrouki2024Stochastic}.
\item We test the simulation results in physical experiments with small-scale vehicles in the Cyber-Physical Mobility Lab (CPM Lab).
\end{enumerate}

\section{Preliminaries}

\subsection{DPC Formulation}

Bold lowercase and uppercase symbols denote vectors and matrices, respectively; scalar quantities are written in lightface. For a signal $\bm{a}$ and a positive-semidefinite matrix \(\bm W\), we define \(\|\bm a\|^2_{\bm W}=\bm a^\mathsf{T} \bm W \bm a\). We denote the value of the signal $\bm a$ at time $k+i$
predicted at time $k$ by $\bm{a}_{i|k}$. We denote a sequence over a horizon $N$ as
$\bm{a}_{\cdot|k} := \operatorname{col}(\bm{a}_{1|k}, \dots, \bm{a}_{N|k})$,
where $\operatorname{col}(\bm{a}_1, \dots, \bm{a}_n) :=
\begin{bmatrix} \bm{a}_1^{\mathsf{T}} & \cdots & \bm{a}_n^{\mathsf{T}} \end{bmatrix}^{\mathsf{T}}$. 

Let
\[
\bm u^{\mathrm d}=\mathrm{col}(\bm u^{\mathrm d}_0,\dots,\bm u^{\mathrm d}_{T-1}), \qquad
\bm y^{\mathrm d}=\mathrm{col}(\bm y^{\mathrm d}_0,\dots,\bm y^{\mathrm d}_{T-1}),
\]
denote the trajectory of recorded control inputs \(\bm u_k^{\mathrm d}\) and outputs \(\bm y_k^{\mathrm d}\) with input and output dimensions \(n_u\) and \(n_y\). We construct input and output Hankel matrices of depth \(T_{\mathrm{ini}}+N\), where \(T_{\mathrm{ini}}\) is the initial trajectory length and \(N\) the prediction horizon. Thus, the columns of each matrix are overlapping trajectories of length \(T_{\mathrm{ini}}+N\). Partitioned into past and future parts, their form is
\[
{\bm H}_{T_{\mathrm{ini}}+N}(\bm u^{\mathrm d})=
\begin{bmatrix}
{\bm U}_{\mathrm p}\\
{\bm U}_{\mathrm f}
\end{bmatrix},
\qquad
{\bm H}_{T_{\mathrm{ini}}+N}(\bm y^{\mathrm d})=
\begin{bmatrix}
{\bm Y}_{\mathrm p}\\
{\bm Y}_{\mathrm f}
\end{bmatrix}.
\]
The blocks with index p and f collect the past and future segments of each trajectory, respectively. 

At sampling instant \(k\), let \(\bm u_{\mathrm{ini}|k}\) and \(\bm y_{\mathrm{ini}|k}\) denote the measured initial trajectory. We write the stacked planned inputs, predicted outputs, and references as
\[
\begin{aligned}
\bm u_{\cdot|k}&=\mathrm{col}(\bm u_{0|k},\dots,\bm u_{N-1|k}),\\
\bm y_{\cdot|k}&=\mathrm{col}(\bm y_{1|k},\dots,\bm y_{N|k}),\\
\bm r_{\cdot|k}&=\mathrm{col}(\bm r_{1|k},\dots,\bm r_{N|k}).
\end{aligned}
\]
The predicted trajectory is parameterized through a coefficient vector \(\bm g_k\). The regularized DPC problem is
\begin{subequations}
\begin{align}
\min_{\substack{\bm g_k,\,\bm\sigma_{u,k},\,\bm\sigma_{y,k},\\
\bm u_{\cdot|k},\,\bm y_{\cdot|k}}}\quad &
\sum_{i=1}^{N} \|\bm y_{i|k}-\bm r_{i|k}\|_{\bm Q}^2
+\sum_{i=0}^{N-1}\|\bm u_{i|k}\|_{\bm R}^2
\nonumber\\
&\quad
+\lambda_g\|\bm g_k\|_2^2+\lambda_\sigma\|\bm\sigma_{y,k}\|_2^2
+\lambda_{\sigma}\|\bm\sigma_{u,k}\|_2^2
\\
\text{such that}\quad &
\begin{bmatrix}
{\bm U}_{\mathrm p}\\
{\bm Y}_{\mathrm p}\\
{\bm U}_{\mathrm f}\\
{\bm Y}_{\mathrm f}
\end{bmatrix}\bm g_k
=
\begin{bmatrix}
\bm u_{\mathrm{ini}|k}+\bm\sigma_{u,k}\\
\bm y_{\mathrm{ini}|k}+\bm\sigma_{y,k}\\
\bm u_{\cdot|k}\\
\bm y_{\cdot|k}
\end{bmatrix},
\quad \bm 1^\mathsf{T} \bm g_k=1,
\\
& \bm u_{i|k}\in\mathcal U,\qquad i=0,\dots,N-1,
\\
& \bm y_{i|k}\in\mathcal Y,\qquad i=1,\dots,N.
\end{align}
\label{eq:DPC}
\end{subequations}
\noindent\text{where} the positive-semidefinite matrices \(\bm Q \in \mathbb S^{n_y}_+\) and \(\bm R \in \mathbb S^{n_u}_+\) weight the tracking and input costs, respectively. The slacks \(\bm\sigma_{u,k}\) and \(\bm\sigma_{y,k}\), weighted by \(\lambda_\sigma>0\), ensure feasibility in the presence of noise and nonlinearity, and \(\lambda_g>0\) weights the regularization. The constraint \(\bm 1^\mathsf{T}\bm g_k=1\) restricts the prediction to affine combinations of recorded trajectories, which represent the affine local dynamics of a nonlinear system around an operating point \cite{berberich2022Linear}. The polyhedral sets \(\mathcal U\subseteq \mathbb R^{n_u}\) and \(\mathcal Y\subseteq \mathbb R^{n_y}\) denote the input and output constraint sets, respectively. The resulting optimization problem is a quadratic program.

\subsection{Contextual Sampling}\label{sec:sampling}

Since the computational complexity of \eqref{eq:DPC} scales with the number of available trajectories, we use contextual sampling to select a subset of trajectories online. Beforehand, we standardize each input and output channel to zero mean and unit standard deviation and align the trajectories to the initial trajectory as in \cite{beerwerth2025Less}. At each sampling instant \(k\), the initial trajectory and the reference are compared with the columns of the past and future Hankel blocks.

For each trajectory \(j\), we compute Euclidean distances for the past data,
\begin{align*}
d^{u}_{\mathrm p,j} &= \|{\bm U}_{\mathrm p,j}-\bm u_{\mathrm{ini}|k}\|_2, &
d^{y}_{\mathrm p,j} &= \|{\bm Y}_{\mathrm p,j}-\bm y_{\mathrm{ini}|k}\|_2,
\end{align*}
and for the future data,
\begin{align*}
d_{\mathrm f,j} &= \|{\bm Y}_{\mathrm f,j}-\bm r_{\cdot|k}\|_2.
\end{align*}
Here, \({\bm U}_{\mathrm p,j}, {\bm Y}_{\mathrm p,j}\), and \({\bm Y}_{\mathrm f,j}\) denote the \(j\)-th columns of the corresponding Hankel blocks.

The input and output distances are then combined into one past score per trajectory,
\[
d_{\mathrm p,j}=d^{u}_{\mathrm p,j}+d^{y}_{\mathrm p,j}.
\]
The past and future distance vectors are normalized to zero mean and unit standard deviation,
\[
\bar d_{\mathrm p,j}=\frac{d_{\mathrm p,j}-m_{\mathrm p}}{s_{\mathrm p}},
\qquad
\bar d_{\mathrm f,j}=\frac{d_{\mathrm f,j}-m_{\mathrm f}}{s_{\mathrm f}},
\]
and combined into the score \(d_j=\bar d_{\mathrm p,j}+\bar d_{\mathrm f,j}\). Here, $m_{\mathrm p}, m_{\mathrm f}$ and $s_{\mathrm p}, s_{\mathrm f}$ denote the mean and standard deviation of $\bm{d}_{\mathrm p}$ and $\bm{d}_{\mathrm f}$, respectively. This normalization prevents either the past or future distance from dominating the score solely because of scale differences.

We select $n$ trajectories in the form of an index set \(\mathcal I\) from the scores \(d_j\) and compile the reduced matrices \(\tilde{\bm U}_{\mathrm p}\in\mathbb R^{T_{\mathrm{ini}}n_u\times n},\tilde{\bm Y}_{\mathrm p}\in\mathbb R^{T_{\mathrm{ini}}n_y\times n},\tilde{\bm U}_{\mathrm f}\in\mathbb R^{Nn_u\times n},\tilde{\bm Y}_{\mathrm f}\in\mathbb R^{Nn_y\times n}\) entering optimization problem \eqref{eq:DPC}. Thus, the controller operates on a subset of trajectories that is relevant to both the initial trajectory and the demanded reference. However, since friction is not measured, contextual sampling still selects trajectories recorded at another friction, when they are close to the initial trajectory and demanded reference.

\section{DPC and Varying Friction}\label{sec:pooling}

To apply DPC to autonomous vehicle racing under varying friction, we pool trajectory data recorded at different friction levels. At each sampling instant, we refer to the data recorded at the current friction level as matched data and the data recorded at other friction levels as mismatched data. Pooled data combine the matched and mismatched data. Accordingly, we define three DPC variants. Pooled DPC uses the pooled data and requires no knowledge of the friction level. Matched DPC uses the matched data. It requires knowing the current friction level and serves as an oracle baseline for the prediction accuracy achievable with the recorded data. Mismatched DPC uses the mismatched data and represents DPC whose data originate from a wrong friction level. We refer to their predictions as the pooled, matched, and mismatched predictions, respectively. 
This section analyzes how matched and mismatched data contribute to the pooled prediction and formulates practical conditions under which pooled DPC approaches the prediction accuracy of matched DPC. To keep the analysis compact, we suppress the sampling-instant index \(k\) in this section. Let the pooled data consist of the matched subset \(\mathcal D_1\) and the mismatched subset \(\mathcal D_2\). We partition the matrices in~\eqref{eq:DPC} and \(\bm g=\mathrm{col}(\bm g_1,\bm g_2)\) accordingly, such that \(\bm y=\tilde{\bm Y}_{\mathrm f,1}\bm g_1+\tilde{\bm Y}_{\mathrm f,2}\bm g_2\). To focus our analysis on the way DPC predicts, we fix the planned input \(\bm u\), assume inactive output constraints and omit the tracking cost. Then~\eqref{eq:DPC} becomes a regularized least-squares problem with a closed-form solution for the coefficient weights \(\bm g\):
\begin{equation}
\begin{aligned}
\min_{\boldsymbol g}\quad & \lambda_g\|\boldsymbol g\|_2^2+\lambda_\sigma\left\|\begin{bmatrix}\tilde{\boldsymbol U}_{\mathrm p}\\ \tilde{\boldsymbol Y}_{\mathrm p}\end{bmatrix}\boldsymbol g-\begin{bmatrix}\boldsymbol u_{\mathrm{ini}}\\ \boldsymbol y_{\mathrm{ini}}\end{bmatrix}\right\|_2^2 \\
\text{such that}\quad & \begin{bmatrix}\tilde{\boldsymbol U}_{\mathrm f}\\ \boldsymbol 1^\mathsf{T}\end{bmatrix}\boldsymbol g=\begin{bmatrix}\boldsymbol u\\ 1\end{bmatrix}.
\end{aligned}
\label{eq:lsp}
\end{equation}
We stack its left-hand sides into $\boldsymbol A$, with columns $\bm a_j$, and its right-hand sides into $\boldsymbol b$,
\[
\boldsymbol A=\begin{bmatrix}\tilde{\boldsymbol U}_{\mathrm p}\\ \tilde{\boldsymbol Y}_{\mathrm p}\\ \tilde{\boldsymbol U}_{\mathrm f}\\ \boldsymbol 1^\mathsf{T}\end{bmatrix},\qquad
\boldsymbol b=\begin{bmatrix}\boldsymbol u_{\mathrm{ini}}\\ \boldsymbol y_{\mathrm{ini}}\\ \boldsymbol u\\ 1\end{bmatrix}.
\]
The closed-form solution of~\eqref{eq:lsp} is $\bm{g} = \bm{A}^\mathsf{T}\bm{M}^{-1}\bm{b}$, with
\begin{equation*}
  \bm{M} = \bm{A}\bm{A}^\mathsf{T} + \frac{\lambda_g}{\lambda_\sigma}
  \begin{bmatrix} \bm{I} & \bm{0} \\ \bm{0} & \bm{0} \end{bmatrix},
\end{equation*}
where $\bm{M}$ is invertible if $\begin{bmatrix}\tilde{\bm{U}}_\mathrm{f}^\mathsf{T} & \bm{1}\end{bmatrix}^\mathsf{T}$ has full row rank.
With $\bm{\nu} := \bm{M}^{-1}\bm{b}$, each recorded trajectory receives the coefficient $g_j = \bm{a}_j^\mathsf{T}\bm{\nu}$, i.e., its alignment with a vector $\bm{\nu}$ that is shared by all trajectories and determined by the initial trajectory and the planned inputs.
Partitioning $\bm{A} = [\bm{A}_1\ \bm{A}_2]$ into the columns corresponding to $\mathcal{D}_1$ and $\mathcal{D}_2$ decomposes the prediction into a matched and a mismatched contribution,
\begin{equation}
  \bm{y} = \tilde{\bm{Y}}_{\mathrm{f},1}\bm{A}_1^\mathsf{T}\bm{\nu} + \tilde{\bm{Y}}_{\mathrm{f},2}\bm{A}_2^\mathsf{T}\bm{\nu}.
\end{equation}
Since~\eqref{eq:lsp} penalizes both the coefficients and the slacks, it tends to assign large coefficients to trajectories that reproduce the initial trajectory and the planned inputs closely.
The share of each subset in $\bm{y}$ therefore depends on how well its trajectories align with $\bm{\nu}$, which depends on the subset's size, scaling, and coverage. We hypothesize that, if fitting the initial trajectory and the planned inputs requires larger coefficients for the mismatched subset than for the matched subset, the pooled prediction relies more strongly on the matched subset and resembles the matched prediction. To what extent the remaining contributions of \(\mathcal D_2\) change the prediction then depends on how clearly the initial trajectory indicates the current friction level, and to what extent the two subsets produce different future outputs under identical planned inputs. We distinguish three cases.

\begin{remark}[Separation\label{rem:separation}]
If the initial trajectory indicates the current friction level, trajectories of \(\mathcal D_2\) align only weakly with it. Under comparable coverage, fitting the initial trajectory and planned inputs with \(\mathcal D_2\) then requires larger coefficients or a larger slack. Both are penalized in \eqref{eq:lsp}. Thus, the pooled prediction is expected to approach the matched prediction. The regularization does not exclude \(\mathcal D_2\), so separation is expected to remain partial.
\end{remark}

\begin{remark}[Benign mixing]\label{rem:mixing}
If both subsets produce the same future outputs for the initial trajectory and planned inputs, mixing trajectories from both subsets does not degrade prediction accuracy. For friction-dependent vehicle dynamics, this approximately describes straight segments and gentle maneuvers in which the tires operate far from their limits.
\end{remark}

\begin{remark}[Misattribution]\label{rem:misattribution}
If both subsets align with the initial trajectory but identical planned inputs lead to different future outputs, mixing trajectories from both subsets degrades prediction accuracy. This can occur at maneuver entry, after a friction change, or when the initial trajectory does not yet contain a maneuver that indicates the current friction. The mismatched prediction is biased wherever the responses under the two friction levels differ.
\end{remark}

The closed-form solution holds for any data, but which subset aligns more strongly with the initial trajectory depends on the data. Smooth dynamics admit a local affine approximation \cite{berberich2022Linear}, and contextual sampling restricts the retained trajectories to a neighborhood of the current operating point \cite{beerwerth2025Less}. Matched trajectories follow approximately the same local dynamics as the initial trajectory. From this interpretation we formulate three practical conditions to explain when and why pooled DPC predicts nearly as accurately as matched DPC:
\begin{itemize}
\item[(A1)] \emph{Coverage:} the matched subset covers the operating points demanded by the reference, and contextual sampling retains data near them.
\item[(A2)] \emph{Comparable coverage:} both subsets are recorded at similar operating points with comparable amounts of data and excitation, so that one subset is not favored merely by imbalanced coverage.
\item[(A3)] \emph{Informative past:} the friction level is approximately constant over the initial trajectory. Where the two friction levels produce different future outputs under the same planned inputs, the initial trajectory indicates the current friction level clearly.
\end{itemize}
Based on these practical conditions, we expect the pooled prediction to remain close to the matched prediction except briefly after maneuver entries or friction changes and when the demanded reference is not covered by the matched subset.

\section{Evaluation}

The evaluations in Sections~\ref{sec:prediction} to~\ref{sec:implied} are carried out in a vehicle simulation with spatially varying friction \cite{okelly2020f1tenth,nagy2023Ensemble}; Section~\ref{sec:hardware} reports experiments on small-scale vehicles with high- and low-friction tire sets. 
We designed the experiments to support A1 and A2. Additionally, we analyze to what extent the resulting differences between the three DPC variants are explained by A3.

In simulation, the vehicle is modeled by a full-scale multi-body model based on \cite{althoff2017CommonRoad}. The controller input \(\bm u_k=(a_k,\delta_k)\) consists of longitudinal acceleration and steering angle, and the measured output \(\bm y_k=(x_k,y_k,\psi_k,v_k)\) consists of vehicle position, heading, and velocity. The racetrack contains multiple friction zones that induce locally varying dynamics along the lap. All simulations use a sampling time of \qty{100}{\milli\second} and a prediction horizon of \(N=10\) steps. 

\subsection{Prediction Accuracy}\label{sec:prediction}

To assess prediction accuracy across frictions, we record trajectories generated using randomly varying steering and speed targets at friction coefficients \(\mu \in \{0.5, 1.1\}\), \qty{1000}{\second} each. These data operate close to the friction limit: \qty{68}{\percent} and \qty{57}{\percent} of the samples exceed \qty{70}{\percent} of the available grip. 
We partition both datasets into \qty{80}{\percent} recorded data to build the DPC predictor and \qty{20}{\percent} validation data for evaluating the predictor. For validation trajectories from $\mu=0.5$, matched DPC uses the recorded data from $\mu=0.5$, mismatched DPC those from $\mu=1.1$, and pooled DPC both. For validation trajectories from $\mu=1.1$, matched DPC uses the recorded data from $\mu=1.1$, mismatched DPC those from $\mu=0.5$, and pooled DPC both. For each validation trajectory, we select \(n=300\) trajectories from the recorded data by contextual sampling to build the reduced Hankel matrices with an initial trajectory length of \(T_{\mathrm{ini}}=10\). We define the prediction error as the $N$-step root-mean-square (RMS) position error, 
\[e_{\mathrm{pred},k}=\sqrt{\frac{1}{N}\sum_{i=1}^{N}\bigl((\hat{x}_{i|k}-x_{k+i})^2+(\hat{y}_{i|k}-y_{k+i})^2\bigr)}\,,\]
with the predicted positions \(\hat{x}_{i|k},\hat{y}_{i|k}\) and the measured positions \(x_{k+i},y_{k+i}\).
The prediction depends on the ratio \(\lambda_\sigma/\lambda_g\); we use \(\lambda_g=5\) and \(\lambda_\sigma=50\). The predictions degrade at extreme ratios: if the regularization dominates (\(\lambda_\sigma/\lambda_g\le 10^{-3}\)), all predictions largely ignore the initial trajectory; if it is negligible (\(\lambda_\sigma/\lambda_g\ge 10^{4}\)), the predictions become ill-conditioned and the pooled prediction no longer favors the matched data.

\begin{figure}[!htb]
    \centering
    \includegraphics[width=\columnwidth]{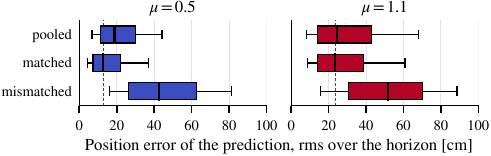}
    \caption{Distribution of the $N$-step RMS prediction error for (1) pooled DPC, (2) matched DPC, and (3) mismatched DPC. Boxes show quartiles and median, whiskers the 10th to 90th percentiles, and the dashed line the median of matched DPC. Colors indicate the evaluation friction.}
    \label{fig:prediction_error}
\end{figure}

Fig.~\ref{fig:prediction_error} summarizes the resulting distributions of the prediction errors. Matched DPC achieves the lowest median prediction errors: \qty{12.7}{\centi\metre} at \(\mu=0.5\) and \qty{23.6}{\centi\metre}  at \(\mu=1.1\). Compared to matched DPC, mismatched DPC increases the median prediction error by \qty{29.8}{\centi\metre} at \(\mu=0.5\) and by \qty{28.1}{\centi\metre} at \(\mu=1.1\). This indicates that trajectories recorded at mismatched friction levels do not represent the dynamics at the matched friction level well. %

Pooled DPC increases the median prediction error by \qty{6.1}{\centi\metre} at \(\mu=0.5\) and by \qty{1.0}{\centi\metre} at \(\mu=1.1\). Thus, pooled DPC approaches the prediction accuracy of matched DPC at \(\mu=1.1\). However, at \(\mu=0.5\) the prediction error of pooled DPC is noticeably higher. This asymmetry is consistent with A3. For initial trajectories from \(\mu=0.5\), data recorded at \(\mu=1.1\) remain competitive in terms of regularization cost. Conversely, for initial trajectories from \(\mu=1.1\), the data recorded at \(\mu=0.5\) incur a higher regularization cost. Fitting the initial trajectory as closely as possible with only the mismatched or only the matched trajectories among the selected ones requires a median coefficient norm \(\|\bm g\|_2\) of 1.09 versus 0.85 at \(\mu=0.5\), and of 2.80 versus 0.89 at \(\mu=1.1\). The initial trajectories from \(\mu=1.1\) therefore indicate their current friction level more clearly.

To check whether the trajectory selection or the regularization favors the matched data, we evaluate the selected trajectories and the solution of~\eqref{eq:lsp} for pooled DPC. Contextual sampling barely separates the subsets: in the median, \qty{55}{\percent} of the selected trajectories are matched at both friction levels. The regularization then shifts the prediction toward the matched subset. We measure the share of the matched subset as \(\|\tilde{\bm Y}_1\bm g_1\|_2^2/(\|\tilde{\bm Y}_1\bm g_1\|_2^2+\|\tilde{\bm Y}_2\bm g_2\|_2^2)\) with \(\tilde{\bm Y}_i=\mathrm{col}(\tilde{\bm Y}_{\mathrm p,i},\tilde{\bm Y}_{\mathrm f,i})\). Its median is 0.83 at \(\mu=0.5\) and 0.92 at \(\mu=1.1\). Thus, the separation results from the regularization, as hypothesized in Section~\ref{sec:pooling}, rather than from the trajectory selection.

\subsection{Trajectory Tracking on a Racetrack with Varying Friction}\label{sec:tracking}

To assess how the prediction accuracy connects to tracking performance, we evaluate the tracking error of the three DPC variants and compare it with the nominal NMPC of the TUM-CONTROL framework \cite{zarrouki2024Stochastic}. Our evaluation racetrack incorporates three friction zones, \(\mu \in \{0.5, 0.8, 1.1\}\). Both the DPC data and the identification maneuvers of the NMPC are collected only at \(\mu=0.5\) and \(\mu=1.1\), so that the zone with \(\mu=0.8\) tests a friction between the recorded ones.

For the NMPC, we use its nominal dynamic vehicle model and provide the ground-truth model parameters from the simulation except for the Pacejka tire parameters \(B_\mathrm{f}\), \(B_\mathrm{r}\), \(D_\mathrm{f}\), and \(D_\mathrm{r}\). These are identified from dedicated excitation maneuvers performed at friction coefficients \(\mu=0.5\) and \(\mu=1.1\). The parameters are obtained by minimizing the $N$-step prediction error of the longitudinal and lateral velocity in the vehicle frame and the yaw rate, \((v_x,v_y,\dot\psi)\), while keeping the remaining model parameters fixed. We tuned three variants of NMPC using the parameter sets identified for \(\mu=0.5\) and \(\mu=1.1\), as well as their arithmetic midpoint. For the evaluation, we use the arithmetic midpoint, since it achieved the lowest tracking error.

For DPC, the data are collected by tracking the raceline with a pure-pursuit controller at different scales of the speed profile at \(\mu=0.5\) and \(\mu=1.1\). Pooled DPC uses the pooled data on the entire racetrack. Matched DPC switches to the data from the current friction level at each transition. Conversely, mismatched DPC switches to the data from the other friction level at each transition. In the \(\mu=0.8\) zone, which neither subset covers, all three DPC controllers use the pooled data. At each sampling instant, contextual sampling selects \(300\) trajectories to build the reduced Hankel matrices. The initial trajectory length is $T_\mathrm{ini} = 5$. Since both subsets are recorded with the same maneuvers, 97 to \qty{99}{\percent} of the operating points of either subset have a counterpart in the other within 0.1 standard deviations of speed and lateral acceleration.

All controllers use the same raceline, speed profile, reference trajectory generation procedure, and sampling time of \qty{100}{\milli\second}. Each controller is evaluated over \(11\) laps for each of three seeds. We discard the first lap to remove initialization effects and aggregate the results of the remaining \(10\) laps.
We define the tracking error as the unsigned cross-track error from the raceline,
\begin{equation}
  e_{\mathrm{track},k} = \min_{p\in[0,L]} \sqrt{\bigl(x_k - x_\mathrm{ref}(p)\bigr)^2 + \bigl(y_k - y_\mathrm{ref}(p)\bigr)^2} ,
\end{equation}
where \(x_\mathrm{ref}(p),y_\mathrm{ref}(p)\) are the reference positions on the raceline at arc length \(p\) on a track of length \(L\). Additionally, we evaluate the one-step prediction error
\[
e_{\mathrm{pred},k}^{(1)}=\sqrt{(\hat x_{1|k}-x_{k+1})^2+(\hat y_{1|k}-y_{k+1})^2}.
\]

\begin{table}[!t]
\caption{Median closed-loop tracking and one-step prediction errors over all steps aggregated for each zone.}
\label{tab:tracking}
\centering
\setlength{\tabcolsep}{2.9pt}
\begin{tabular}{l cccc cccc}
\toprule
 & \multicolumn{4}{c}{Tracking error [\unit{\centi\metre}]} & \multicolumn{4}{c}{Prediction error [\unit{\milli\metre}]} \\
 & \multicolumn{3}{c}{zone $\mu$} & & \multicolumn{3}{c}{zone $\mu$} & \\
\cmidrule(lr){2-4}\cmidrule(lr){6-8}
 & 0.5 & 0.8 & 1.1 & lap & 0.5 & 0.8 & 1.1 & lap\\
\midrule
Pooled DPC & 12.1 & 15.4 & 14.0 & 13.7 & 2.1 & 2.6 & 1.8 & 2.0 \\
Matched DPC & 14.5 & 16.0 & 13.0 & 14.4 & \textbf{1.5} & 2.6 & \textbf{0.7} & \textbf{1.5}\\
Mismatched DPC & \textbf{11.6} & 14.8 & 19.2 & 15.0 & 4.6 & 2.6 & 3.7 & 3.8 \\
NMPC & 17.4 & \textbf{12.1} & \textbf{9.7} & \textbf{13.3} & -- & -- & -- & -- \\
\bottomrule
\end{tabular}
\end{table}

\begin{figure}[!htb]
    \centering
    \includegraphics[width=\columnwidth]{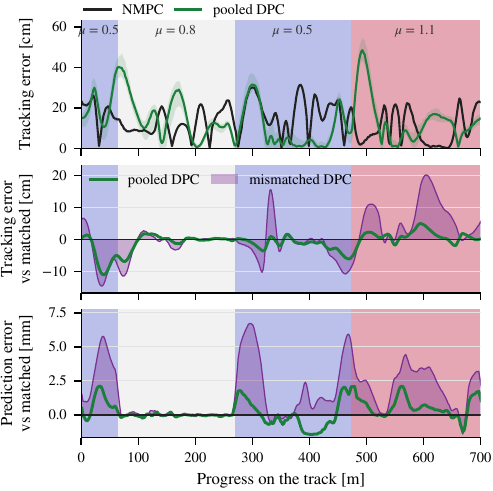}
    \caption{Top: tracking error over racetrack progress for pooled DPC and NMPC. The curves show the median over 30 evaluated laps, and the shaded regions indicate the corresponding minimum and maximum errors. Middle and bottom: differences in tracking and prediction errors between pooled DPC or mismatched DPC and matched DPC. The background color indicates the track friction profile, with blue denoting low friction (\(\mu=0.5\)), gray intermediate friction (\(\mu=0.8\)), and red high friction (\(\mu=1.1\)).}
    \label{fig:tracking_error}
\end{figure}
Table~\ref{tab:tracking} lists the tracking error and the one-step prediction error per friction zone and Fig.~\ref{fig:tracking_error} shows the resulting tracking and one-step prediction error along the racetrack. Over the full lap, pooled DPC tracks the raceline with a tracking error of \qty{13.7}{\centi\metre} and NMPC with a tracking error of \qty{13.3}{\centi\metre}. Across the three seeds, the lap tracking error of each controller varies by at most \qty{0.2}{\centi\metre}. Their tracking errors differ by zone: NMPC tracks worse at \(\mu=0.5\), while pooled DPC tracks worse at \(\mu=1.1\). In the intermediate zone, the median tracking errors of both controllers lie within \qty{3.3}{\centi\metre} of each other.

The middle and bottom panels of Fig.~\ref{fig:tracking_error} compare the tracking error and one-step prediction error of the three DPC variants. Overall, pooled DPC achieves the lowest median tracking error. However, in terms of prediction accuracy, matched DPC achieves the lowest median one-step prediction error. Thus, the tracking performance does not reflect the prediction accuracy of the three DPC variants. 

In the friction zones covered by the recorded data, matched DPC achieves the lowest median one-step prediction error: \qty{1.5}{\milli\metre} at \(\mu=0.5\) and \qty{0.7}{\milli\metre}  at \(\mu=1.1\). Compared to matched DPC, mismatched DPC increases the prediction error by \qty{3.1}{\milli\metre} at \(\mu=0.5\) and by \qty{3.0}{\milli\metre} at \(\mu=1.1\). Compared to matched DPC, pooled DPC increases the prediction error by \qty{0.6}{\milli\metre} at \(\mu=0.5\) and \qty{1.1}{\milli\metre} at \(\mu=1.1\). Thus, the pooled predictions are closer to the matched ones than they are to the mismatched ones. 
The pooled prediction is not uniformly less accurate than the matched prediction. The largest deviations occur at maneuver entries and friction transitions. This is consistent with A3 and Case~\ref{rem:misattribution}: the future vehicle motion depends on the current friction, while the initial trajectory does not indicate it yet.

Lower prediction error does not necessarily yield lower tracking error. At \(\mu=0.5\) mismatched DPC yields the lowest tracking error despite having the highest prediction error. All three DPC variants cut the two sharp corners. Mismatched DPC overestimates the available grip and predicts the vehicle to move farther inside than it does. Consequently, mismatched DPC cuts the corner less, resulting in an unintended decrease of the tracking error. Conversely, at \(\mu=1.1\), mismatched DPC increases the median tracking error since it underestimates the available grip. Pooled DPC retains most of the tracking performance of matched DPC, increasing the median tracking error by only \qty{1.0}{\centi\metre} (+\qty{7.69}{\percent}).

The computation time of DPC is well within the sampling time of \qty{100}{\milli\second}. One control step, including contextual sampling and the solution of~\eqref{eq:DPC}, takes \qty{21.5}{\milli\second} in the median and \qty{44.5}{\milli\second} at the 95th percentile on an Apple M3 Pro CPU.

To test A1, we apply pooled DPC to the unseen Nürburgring racetrack with friction zones within the recorded range. With the reference speed capped at \qty{20}{\metre\per\second} as on the evaluation racetrack, pooled DPC completes the laps with a tracking error of \qty{8.8}{\centi\metre}, compared to \qty{6.7}{\centi\metre} for NMPC; the largest deviation occurs at the hairpin, whose curvature is twice the largest curvature in the recorded data. Without the cap, \qty{68}{\percent} of the speed profile exceeds the speeds covered by the data, and pooled DPC fails. Transfer to a new racetrack thus requires that the data cover the demanded speeds and curvatures (A1).

\subsection{Friction Implied by the Pooled Prediction}\label{sec:implied}

In this section, we investigate to what extent the pooled prediction approaches the matched prediction along the racetrack. We expect the regularization of pooled DPC to favor matched data wherever the initial trajectory indicates the current friction.

As in Section~\ref{sec:pooling}, we fix the inputs planned by pooled DPC and solve \eqref{eq:lsp} with the pooled data and with the two friction-specific datasets: data from \(\mu_\mathrm{low}=0.5\) and data from \(\mu_\mathrm{high}=1.1\). We approximate the pooled prediction \(\bm y_{\cdot|k}\) as a convex combination of the two friction-specific predictions and determine the interpolation weight,
\begin{equation}
\begin{aligned}
w_k = \operatorname*{arg\,min}_{w\in[0,1]}\bigl\|\bm y_{\cdot|k}-w\,\bm y_{\cdot|k}^{(\mathrm{low})}-(1-w)\,\bm y_{\cdot|k}^{(\mathrm{high})}\bigr\|_2,
\end{aligned}
\end{equation}
where \(w_k\) denotes the interpolation weight at sampling instant \(k\).
We define the implied friction as the weighted average
\begin{equation}
\mu_{\mathrm{impl},k}
=
w_k\mu_\mathrm{low} + (1-w_k)\mu_\mathrm{high}.
\end{equation}
The implied friction indicates which of the two friction-specific predictions the pooled prediction resembles more closely.
Since \(w_k\) is not identifiable where the two predictions coincide, we weight each sample by \(\|\bm y_{\cdot|k}^{(\mathrm{low})}-\bm y_{\cdot|k}^{(\mathrm{high})}\|_2^2\) when aggregating the implied friction over racetrack segments.

\begin{figure}[!htb]
    \centering
    \includegraphics[width=\columnwidth]{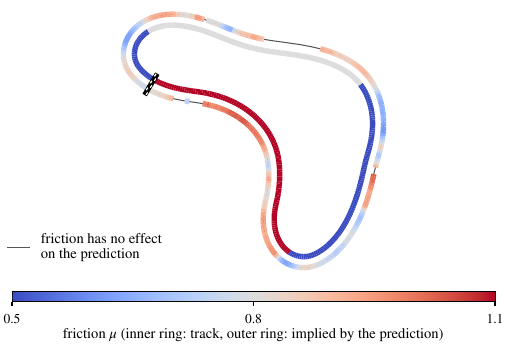}
    \caption{True friction (inner ring) and the friction implied by the pooled prediction (outer ring) aggregated over \qty{10}{\metre} sections of the racetrack. The thin outline marks sections on which the predictions from the data recorded at each friction differ by less than \qty{2}{\centi\metre}.}
    \label{fig:implied_friction}
\end{figure}

Fig.~\ref{fig:implied_friction} compares the true racetrack friction with the implied friction. In the corners, the implied friction follows the spatial variation of the racetrack friction, indicating that the pooled predictions more closely resemble the matched predictions (Case~\ref{rem:separation}). On straights and in gentle bends, it stays near the middle of the range or is not identifiable (Case~\ref{rem:mixing}). The separation is partial: on the segments of the racetrack where the two predictions differ most, the implied friction is 0.77 in the low- and 0.96 in the high-friction zone, respectively. The separation is weaker in the low-friction zone, consistent with the asymmetry in Section~\ref{sec:prediction}: data recorded at \(\mu=1.1\) remain competitive for initial trajectories from \(\mu=0.5\). The vehicle uses \qty{68}{\percent}, \qty{39}{\percent}, and \qty{33}{\percent} of the available grip in the median and \qty{83}{\percent}, \qty{49}{\percent}, and \qty{38}{\percent} at the 95th percentile at \(\mu=0.5\), \(0.8\), and \(1.1\), respectively. This is consistent with Cases~\ref{rem:separation} and~\ref{rem:mixing} and (A3): the separation appears where the friction is indicated by the initial trajectory and affects the vehicle motion.

\subsection{Hardware Experiments}\label{sec:hardware}

We test to what extent the results of Sections~\ref{sec:prediction} to~\ref{sec:implied} hold on physical vehicles, using 1:18-scale vehicles in the CPM Lab \cite{kloock2021Cyberphysical}, which localizes the vehicles with a camera-based indoor positioning system \cite{kloock2020Visionbased}. The input \(\bm u_k\) consists of motor throttle and steering angle, and the output \(\bm y_k\) consists of position, heading, and velocity. Two tire sets represent two friction levels: rubber tires for high friction and the same tires covered with a smooth film for low friction. For each tire set, a pure-pursuit controller tracks a \qty{12}{\metre} raceline at seven scalings of the speed profile for \qty{100}{\second} each. DPC runs on a workstation with a sampling time of \qty{20}{\milli\second}, \(T_{\mathrm{ini}}=10\), \(N=20\). At each sampling instant, contextual sampling selects \(n=100\) trajectories to build the reduced Hankel matrices. We evaluate pooled DPC, matched DPC, and mismatched DPC in three runs of five laps per tire set. Matched DPC uses the data recorded with the same tire set and mismatched DPC the data recorded with the other tire set. Table~\ref{tab:hardware} aggregates the median tracking and one-step prediction error across the runs.

\begin{table}[!t]
\caption{Median tracking and one-step prediction errors over three runs, and mean lap time in small-scale vehicle experiments in the CPM Lab.}
\label{tab:hardware}
\centering
\footnotesize
\setlength{\tabcolsep}{2.8pt}
\begin{tabular}{l ccc ccc}
\toprule
 & \multicolumn{3}{c}{Low-friction tires} & \multicolumn{3}{c}{High-friction tires} \\
\cmidrule(lr){2-4}\cmidrule(lr){5-7}
 & tracking & pred. & lap & tracking & pred. & lap \\
 & [\unit{\centi\metre}] & [\unit{\milli\metre}] & [\unit{\second}] & [\unit{\centi\metre}] & [\unit{\milli\metre}] & [\unit{\second}] \\
\midrule
Pooled DPC & 1.6 & 1.72 & 9.62 & 0.9 & 1.52 & 10.17 \\
Matched DPC & \textbf{1.5} & \textbf{1.69} & 9.51 & 1.1 & 1.56 & \textbf{9.84} \\
Mismatched DPC & 3.0 & 2.08 & \textbf{9.30} & \textbf{0.7} & \textbf{1.47} & 10.35 \\
\bottomrule
\end{tabular}
\end{table}

\begin{figure}[!t]
    \centering
    \includegraphics[width=\columnwidth]{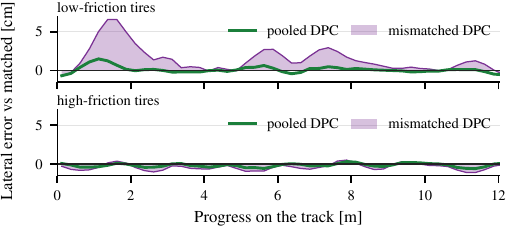}
    \caption{Difference in tracking error between pooled DPC or mismatched DPC and matched DPC in small-scale vehicle experiments in the CPM Lab.}
    \label{fig:hardware}
\end{figure}

Matched DPC achieves a median tracking error of \qty{1.5}{\centi\metre} and a median one-step prediction error of \qty{1.69}{\milli\metre} on low-friction tires and errors of \qty{1.1}{\centi\metre} and \qty{1.56}{\milli\metre} on high-friction tires.
Pooled DPC performs close to matched DPC on both tire sets. The tracking error differs by at most \qty{0.2}{\centi\metre} and the prediction error by at most \qty{0.04}{\milli\metre}. Across the three runs, the median tracking errors of pooled DPC overlap with those of matched DPC on both tire sets, whereas those of mismatched DPC lie outside the range of matched DPC in every run. The noise of the indoor positioning system, estimated from the high-frequency residual of the logged positions, is below \qty{1}{\milli\metre}. With mismatched DPC the results of the two tire sets differ strongly.
On the low-friction tires, mismatched DPC overestimates the available grip and increases the median tracking error by \qty{1.5}{\centi\metre}. On the high-friction tires, all three variants track the corners comparably; the differences in tracking error arise on the low-curvature segments, where the friction hardly affects the vehicle motion (Case~\ref{rem:mixing}). There, mismatched DPC also benefits from driving about \qty{5}{\percent} slower than matched DPC, so its lower tracking error does not indicate an advantage of using mismatched data. Likewise, pooled DPC drives about \qty{3}{\percent} slower than matched DPC on the high-friction tires, which contributes to its lower tracking error.
The computation time of DPC is within the sampling time of \qty{20}{\milli\second}. One control step takes \qty{13.1}{\milli\second} in the median and \qty{15.8}{\milli\second} at the 95th percentile on our workstation.

\section{Conclusion}
This paper investigated to what extent and under which conditions DPC with data pooled from trajectories recorded at varying friction levels retains the prediction accuracy of DPC with data from the current friction (matched data).
In simulation, pooled data keep the multi-step root-mean-square prediction error within 1.5 times that of matched data, whereas mismatched data, i.e., data from a different friction, increase it by a factor of 2.2 to 3.4. In small-scale vehicle experiments in the CPM Lab, pooled and matched data yield tracking errors within \qty{0.2}{\centi\metre} on both the high- and low-friction tires. With mismatched data the tracking error increases on the low-friction tires; on the high-friction tires, the tracking error decreases, but lap time increases by \qty{5}{\percent}. The friction implied by the pooled prediction follows the track friction in the corners, where the friction affects the vehicle motion, and the separation remains partial after maneuver entries and friction changes. DPC thus provides partial implicit adaptation without estimating friction explicitly.

Future work should formalize when the initial trajectory indicates the current friction sufficiently and investigate improved trajectory selection, subset-aware regularization, and data collection for comparable coverage.

\section*{Acknowledgment}
The authors used Claude Opus 5.5 (Anthropic) and GPT-5.6 Sol (OpenAI) to implement parts of the experimental code and to refine parts of the language and exposition of the manuscript. The authors reviewed all AI-generated output and take full responsibility for the content of this paper.

\bibliographystyle{IEEEtran}
\bibliography{root}

\end{document}